# Confronting Interatomic Force Measurements

Omur E Dagdeviren[1,2,*]

[1] Department of Physics, McGill University, Montreal, Quebec, Canada, H3A 2T8
[2] Department of Mechanical Engineering and Materials Science, Yale University, New Haven, Connecticut 06520, USA

*Corresponding author's email: omur.dagdeviren@mcgill.ca

The quantitative interatomic force measurements open a new pathway to materials characterization, surface science, and chemistry by elucidating the force between "two" interacting atoms as a function of their separation. Atomic force microscope is the ideal platform to gauge interatomic forces between the tip and the sample. For such quantitative measurements, either the oscillation frequency or the oscillation amplitude and the phase of a vibrating cantilever are recorded as a function of the tip-sample separation. These experimental measures are subsequently converted into the interatomic force laws. Recently, it has been shown that the most commonly applied mathematical conversion techniques may suffer a significant deviation from the actual force laws. To avoid assessment of unphysical interatomic forces, either the use of very small (i.e., a few picometers) or very large oscillation amplitudes (i.e., a few nanometers) has been proposed. However, the use of marginal oscillation amplitudes gives rise to another problem as it lacks the feasibility due to the adverse signal to noise ratios. Here we show a new mathematical conversion principle that confronts interatomic force measurements while preserving the oscillation amplitude within the experimentally achievable and favorable limits, i.e. tens of picometers. We anticipate that our findings will be the nucleus of reliable evaluation of material properties with a more accurate measurement of interatomic force laws.

Mapping interactions between a sharp probe tip and a sample surface has become possible with the advances of scanning probe microscopy[1]. Among other interactions that can be gauged by employing scanning-probe-based techniques, force measurements are among the most popular choice[2-7]. Material characterization with force measurements dates back to the premier realizations of atomic force microscopes[8]. Initially, forces were recovered by gauging the deflection of the cantilever beam and with the knowledge of cantilever's spring constant[9,10]. Although this direct deflection measurement-based technique is still applied, the degree of locality is dictated by the dimensions of the tip's apex and the size of the contact area. Besides, this deflection-based technique also suffers from the mechanical instabilities that arise within the proximity of the sample's surface. Sudden instabilities (i.e., jumps-in) occur at the exact distance where the gradient of the attractive surface forces becomes larger than the cantilever's spring constant[9,10]. Utilization of cantilevers that feature spring constants much higher than the largest force gradient experienced during the approach has become customary particularly for vacuum applications. While the use of stiff cantilevers eliminates mechanical instabilities, it also impedes the applicability of Hooke's law. The high spring constants reduce the cantilever deflections to values that are too small to resolve with sufficient accuracy using the current sensors.

As an alternative, "dynamic" operational methodologies for atomic force microscopy (AFM) can be applied to assess interatomic forces[11,12]. As Figure 1 illustrates, a probe oscillates in the close vicinity of the surface. The perturbations from the harmonic oscillation of the cantilever due to the tip-sample interaction are measured as a function of the tip-sample separation. To this end, two modulation techniques are most commonly used. The frequency modulation (FM) technique is widely employed, which demodulates the resonance frequency $\Delta f$ of the cantilever beam under the influence of the tip-sample interaction while keeping the oscillation amplitude constant[13]. Alternatively, the amplitude modulation (AM)-based techniques track the change of the oscillation amplitude (*A*) and/or the phase difference between the oscillation and excitation ($\varphi$) while driving with a constant excitation signal[11].

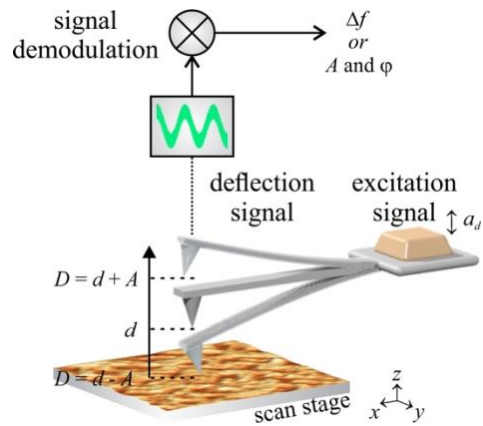

**Figure 1:** Explanation of experimental procedure of interatomic force measurements. The base of the vibrating cantilever is excited with a sinusoidal signal with an oscillation amplitude $a_d$. The cantilever oscillates with an oscillation amplitude, *A*. The tip's apex is at a distance, *d*, away from the surface when it is undeflected. The distance, $D = d - A$, distinguishes itself from the undeflected cantilever position. The deflection signal is demodulated. The resonance frequency shift ($\Delta f$) or the oscillation amplitude (A) and the phase of the cantilever signal ($\varphi$) are recorded as a function of tip-sample separation.

The interatomic forces can be recovered by employing the techniques that were developed around the millennium from the demodulated oscillation signal[14-18]. Equation 1 shows the general form of the conversion technique to assess interatomic forces, $F_{IA}$, both for AM and FM-type force spectroscopy experiments[17,19,20]:

$$F_{IA}(D) = \frac{\partial}{\partial D}\left[2c_z \int_D^\infty PF\left[\textcolor{red}{(z-D)} + \textcolor{orange}{(\sqrt{\frac{A}{16\pi}}\sqrt{z-D})} + \textcolor{green}{(\frac{A^{3/2}}{\sqrt{2(z-D)}})}\right]dz\right] \quad (1)$$

In equation 1, $c_z$ is the spring constant of the cantilever. The first term in the integral, $PF$, is a pre-factor which changes for AM and FM-based force spectroscopy experiments[17,19,20]: The position of the tip is presented by $z$, while $D$ is the nearest tip-sample distance and $A$ is the oscillation amplitude as explained in Figure 1. Different terms in the parenthesis dominate the integral as a function of $A$ and the decay length of the interatomic force, $l$[17,21]. When $A < l$, the first term in the parenthesis, which the term is written in red, dominates the integral. The integral is dominated by the term written in green for $A > l$. Subsequently, the second term (written in orange) of equation 1, which is an approximation term, dominates the integral when $A \approx l$. Recently, it has been shown that the interatomic force measurements that employs equation 1 may deviate significantly from the actual force laws[21,22]. More specifically, sudden changes in the interatomic force laws may impede the reconstruction of the force from the measured quantities due to the limitations of renormalization schemes and approximations applied in equation 1. It was proposed that the oscillation amplitude should be adjusted depending on the distance and the degree of the sudden change in the interatomic forces to avoid problems associated with the limitations of the mathematical conversion principles[21,22]; however, marginal oscillation amplitudes provoke another experimental problem which is the attenuated signal to noise ratio[11,12]. Our numerical analysis reveals that the mathematical instabilities may also arise because of the small amplitude term of equation 1 (i.e. term written in red) in the existence of a sudden change in the interatomic forces as a function of distance. Here we show a new mathematical conversion principle that confronts the reconstruction of interatomic force laws with a significant relaxation of applicable oscillation amplitude range for a successful measurement with the experimentally achievable range of oscillation amplitudes.

Lee et al. developed a general theory for the reconstruction of interatomic force laws for AM-based force spectroscopy experiments, which works for small-amplitude range without any instabilities[18]. However, at large oscillation amplitudes, the technique is mathematically intense and may deviate from the actual force laws[18]. We propose a hybrid technique that employs the general theory for the reconstruction of interatomic force laws to avoid mathematical instabilities of the small amplitude term in equation 1. The hybrid technique for the reconstruction of the interatomic force laws is presented by equation 2:

$$F_{\text{IA}}(D) = \int_D^\infty dz \left[\frac{c_z A_0 \sin \varphi(z)}{Q A(z)} - (c_z - mw^2)\right] + \frac{\partial}{\partial D}\left|2c_z \int_D^\infty PF\left[(\sqrt{\frac{A}{16\pi}}\sqrt{z-D}) + (\frac{A^{3/2}}{\sqrt{2(z-D)}})\right]dz\right| \quad (2)$$

In this hybrid technique, the effective mass, $m$, is replaced by $c_z/(2\pi f_0)^2$. The angular frequency term, $w$, equals to $(2\pi f)$. The term $f$ equals to drive frequency of the cantilever for AM-modulation. The technique can be extended for FM-modulation by introducing ($f_0 + \Delta f$) for $f$. Also, the sinusoidal term in the summation drops for FM-modulation as the phase is kept 90° by the control electronics for a constant oscillation amplitude[12]. For this reason, FM modulation-based force spectroscopy using equation 2 is immune to uncertainty of the oscillation amplitude for small oscillation amplitudes which is a major advantage over the existing techniques[22].

To test our methodology, we followed a commonly used approach for dynamic AFM. We solved the equation of motion of a damped harmonic oscillator with external excitation and a model interatomic force[23,24]. We calculated $A$ and $\varphi$ for AM-modulation and $\Delta f$ for FM-modulation as a function of tip-sample separation. We then reconstructed the model interatomic force field both with equation 1 (i.e., method 1) and equation 2 (i.e., method 2). Figure 2 shows a summary of our results for four different interatomic force models. As Figure 2 illustrates, we can successfully reconstruct the force even after the sudden change in the interatomic force as a function distance, also known as the inflection point[22]. According to former proposals, the lower limit of the required oscillation amplitude approximates to zero for a successful force reconstruction by using method 1 as the derivative of the interatomic force model is discontinuous. As illustrated in Figure 2 a-d, method 1 suffers from mathematical instabilities after the inflection point both for AM and FM modulation-based reconstruction. Method 2, however, can successfully reconstruct the interatomic force model even in the existence of multiple inflection points while experimentally feasible oscillation amplitudes are employed. This results in a significant relaxation of the applicable oscillation amplitude range for force measurements.

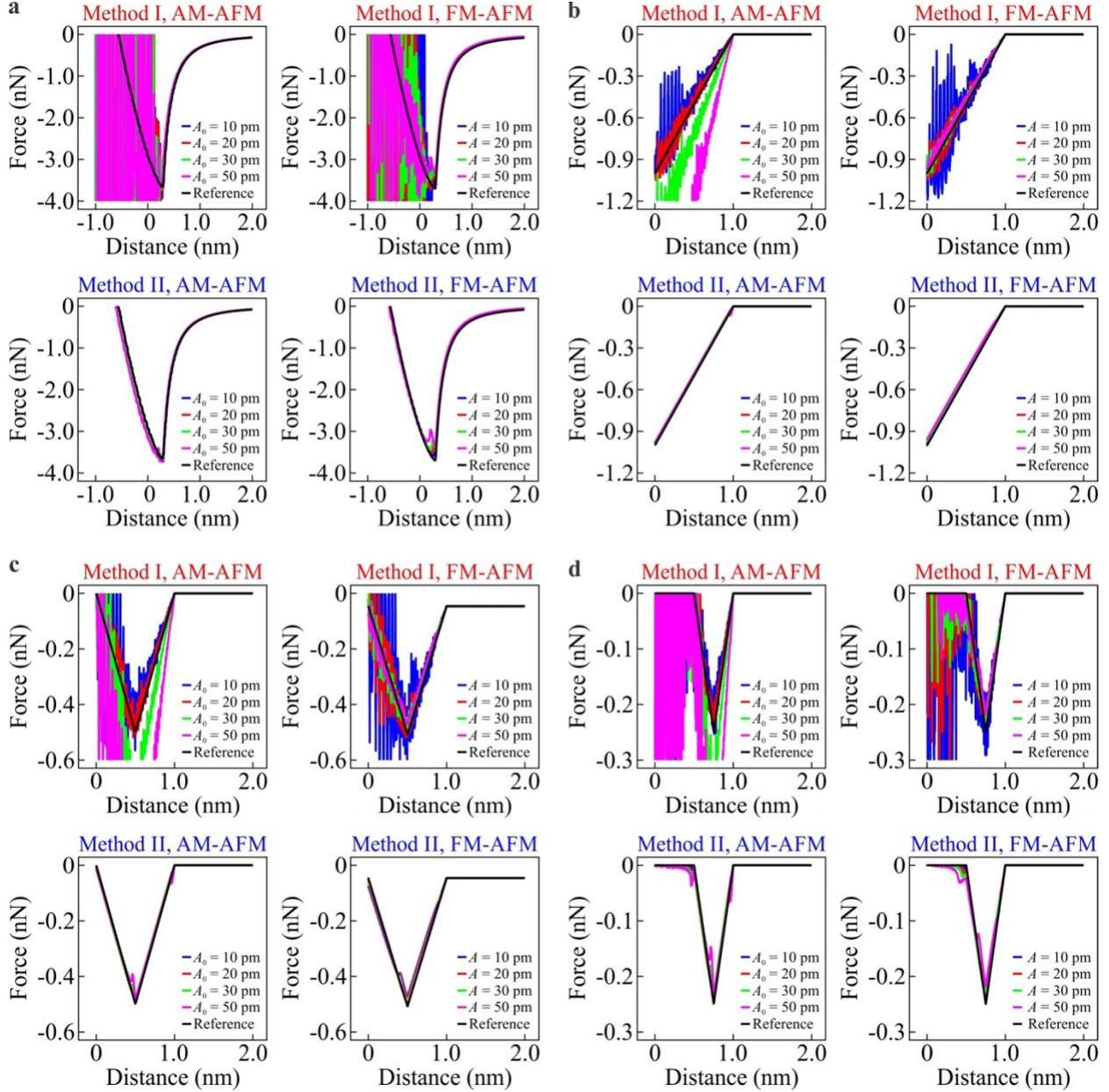

**Figure 2:** The comparison of two different force reconstruction methodologies for different interatomic force models, (a) Hertz-Offset model (for details see Ref [21]), (b) ramp force model, (c) triangle force model with two inflection points, (d) triangle force model with three inflection points. The reconstruction of tip-sample interaction force is presented with different colors for different oscillation amplitudes while the model force is presented in black. (a-d) Our numerical results show that method 1 suffers from mathematical instabilities. (a-d) Method 2 successfully eliminates mathematical instabilities even in the existence of multiple inflection points for oscillation amplitudes that are feasible from the experimental perspective both for AM and FM modulation-based interatomic force reconstruction.

Our technique enables confronting the interatomic force measurements by eliminating mathematical instabilities for experimentally achievable oscillation amplitudes. For this reason, it has a major advantage over existing practices. Although our methodology is significantly more robust than the existing techniques, a more generalized mathematical conversion principle for all oscillation amplitude ranges and all possible interatomic force laws remains an open question.

**Acknowledgements**

I would like to thank Prof. Peter Grütter and Prof. Udo D. Schwarz for fruitful discussions. Financial support from the Natural Sciences and Engineering Research Council of Canada and Le Fonds de Recherche du Québec - Nature et Technologies are gratefully acknowledged.